# Anomalous quasiparticles in the zone center electron pocket of the kagomé ferromagnet Fe$_3$Sn$_2$

Sandy Adhitia Ekahana[1], Y. Soh[1*], Anna Tamai[2], Daniel Gosálbez-Martínez[4,5], Mengyu Yao[3], Andrew Hunter[2], Wenhui Fan[6,7], Yihao Wang[8], Junbo Li[8], Armin Kleibert[1], C. A. F. Vaz[1], Junzhang Ma[9,1], Yimin Xiong[8], Oleg V. Yazyev[4,5], Felix Baumberger[2,1], Ming Shi[1], Gabriel Aeppli[1,4,10,11]

1 Paul Scherrer Institute, Forschungstrasse 111, CH-5232, Villigen, Switzerland

2 Department of Quantum Matter Physics, University of Geneva, 24 Quai Ernest-Ansermet, CH-1211, Geneva, Switzerland

3 Max Planck Institute for Chemical Physics of Solids, D-01187 Dresden, Germany

4 Institut de Physique, Ecole Polytechnique Fédérale de Lausanne (EPFL), CH-1015 Lausanne, Switzerland

5 National Centre for Computational Design and Discovery of Novel Materials MARVEL, Ecole Polytechnique Fédérale de Lausanne (EPFL), CH-1015 Lausanne, Switzerland

6 Beijing National Laboratory for Condensed Matter Physics and Institute of Physics, Chinese Academy of Sciences, P. O. Box 603, Beijing, 100190, China.

7 University of Chinese Academy of Sciences, Beijing 101408, China.

8 Anhui Province Key Laboratory of Condensed Matter Physics at Extreme Conditions, High Magnetic Field Laboratory of the Chinese Academy of Sciences, Hefei 230031, China

9 Department of Physics, City University of Hong Kong, Kowloon, Hong Kong, China

10 Department of Physics, Eidgenössische Technische Hochschule Zurich (ETHZ), CH-8093, Switzerland

11 Quantum Center, Eidgenössische Technische Hochschule Zurich (ETHZ), CH-8093 Zurich, Switzerland

*To whom correspondence should be addressed. Email: yona.soh@psi.ch

**Abstract:**

**The kagomé lattice is a triangular lattice with all corner sites removed on a 2 × 2 superlattice. The enlarged unit cell enhances the possibility of topological effects on the properties of materials constructed from such lattices. One material containing kagome bilayers and featuring both exceptional magnetism and electron transport is the ferromagnetic metal $Fe_3Sn_2$ [1-6]. Notwithstanding the widespread interest in $Fe_3Sn_2$, crystal twinning, difficulties of distinguishing surface from bulk states, and a large unit cell have until now prevented the synchrotron-based spectroscopic observation of sharply resolved quasiparticles near the Fermi surface which could be responsible for the anomalous properties appearing at low temperatures for the material. Here we report microfocused laser-based angle-resolved photoemission spectroscopy (μ-ARPES), which offers the first look at such quasiparticles. The high spatial resolution allows individual crystal twin domains to be examined in isolation, resulting in the discovery of three-fold symmetric electron pockets at the Brillouin zone (BZ) center, not predicted by early tight-binding descriptions but in agreement with density functional theory (DFT) calculations, which also feature Weyl nodes [4]. The quasiparticles in these pockets have a remarkably long mean free paths, and their Fermi surface area is consistent with reported quantum oscillations [7]. At the same time, though, the best-defined Fermi surface is reduced at low temperature, and the quasiparticles generally are marginal in the sense that their wavelength uncertainty is of order the deviation of the quasiparticle wavelength from the Fermi vector. We attribute these manifestations of strong electron-electron interactions to a flat band predicted by our DFT to lie just above the dispersive bands seen in this experiment. Thus, beyond demonstrating the impact of twin averaging for ARPES measurements of band structures, our experiments reveal many-body physics unaccounted for by current theories of metallic kagomé ferromagnets.**

The kagomé lattice was first introduced in 1951 [8] as a natural expansion of square [9], honeycomb, and triangular lattices [10,11] for the study of Ising magnets. An important observation of the original publication was that for antiferromagnetic interactions, no conventional classical order is expected at finite temperatures. Long after its introduction, it remained the province of theory [12,13] until its first experimental discoveries in jarosites [14] and magnetoplumbites [15,16] . A second strand of research concerns the band structure for electrons hosted by such lattices which, in the simplest tight-binding limit, not only share the Dirac crossings of honeycomb lattices such as graphene, but also display flat bands. Particularly noteworthy is the prediction from 2011 based on a tight binding model that a kagomé material which is both ferromagnetic and metallic could display interesting collective phenomena such as the fractional quantum Hall effect in the absence of an external magnetic field [3]. Rhombohedral (space group R-3m, SG-166) $Fe_3Sn_2$ is precisely such a substance[17]. Its building blocks are bilayers of distorted, three-fold (not six-fold) symmetric kagomé layers of Fe atoms alternating with single layers of stanene, as shown in Fig. 1a. Of long-standing interest have been the ferromagnetic properties, including a high Curie temperature of 640 K [1] and a spin reorientation transition near 120 K, which was recently confirmed to be of first order [5,6]. The latter follows from a crossing of different free energies associated with subtly different band structures and resulting Fermi surfaces for the different moment directions [4,6] (see figure SI8 in supplementary information (SI)). The electronic band structure itself has been directly investigated by photoemission [4,18-20], scanning tunneling microscopy [21], and bulk transport measurements [2,5,22-24]. The simplest tight-binding calculations (single orbital per site of kagomé lattice) for idealized two-dimensional slabs reveal flat bands and Dirac points, and there are claims of their observation in angle-resolved photoemission (ARPES), bulk Hall effect, and de Haas-van Alphen measurements [7,18]. In contrast, density functional theory (DFT) for the bulk material yields neither obvious flat bands nor Dirac nodes (see figure SI10) but does reveal Weyl points near the Fermi surface [4]. Nonetheless, it accounts for the seeming appearance of Dirac nodes in ARPES in terms of superposed bulk and surface bands [4,20]. The ARPES data collected so far do not display sharply resolved Weyl nodes. However, there is another feature that unambiguously distinguishes the tight-binding calculations from DFT predictions, namely, only DFT predicts electron pockets surrounding the zone centers (see figure 1c and figure SI7 – SI9[4]). Furthermore, as shown in figure 1c, the pockets are highly sensitive to electron correlation effects as parametrized by the effective Coulomb interaction $U$ in DFT + U calculations, and so could be exploited for measurement of $U$. We have therefore set out to search for and characterize these pockets experimentally.

Here, we report the first detailed experimental study of the distinct electron pockets, present in lower resolution and more surface-sensitive measurements at the Brillouin zone (BZ) center of $Fe_3Sn_2$ [4], by employing state-of-the-art micro-focused laser ARPES (μ-ARPES) to overcome the presence of crystallographic twinning. Our results are consistent with the same density functional calculations which suggest that the Dirac-like cones reported in conventional synchrotron-based ARPES [4,18] derive from a coincidence between minima and maxima of surface and bulk bands [4,20], respectively. At the lowest temperature (6 K), there are very sharp quasiparticles which define small Fermi surfaces around the Brillouin zone centre, and their inverse lifetimes show the rapid linear increase with binding energy characteristic of strongly interacting fermions. On warming, the most sharply resolved band crossing the Fermi surface effectively dissolves, possibly due to hybridization and subsequent localization involving a flat band placed by DFT just above the chemical potential.

**Motivation for spatially resolved ARPES experiments**

ARPES is the natural technique for determining the electronic band structure and has been applied on numerous occasions to Fe$_3$Sn$_2$ [4,18-20]. However, there are four potential difficulties in assessing ARPES data, beyond the usual challenge of distinguishing surface from bulk contributions. A first challenge relates to the large number of atoms per unit cell (18 Fe and 12 Sn), leading to closely spaced bands which, due to magnetic order, will also be spin-split. A second difficulty is the possibility of crystal twinning originating during the crystal growth and which corresponds to rotation by $\delta = m \times 60°, m = \pm(1,3,5)$ (figure 1a). A third problem is the possibility of different surface terminations, which would yield different ARPES results. A fourth aspect is the presence of ferromagnetism where, due to spin-orbit coupling, different magnetic domains are associated with different electronic structures [4].

The first problem can only be remedied by ultra-high-resolution spectroscopy, while the remaining three are all linked to different types of multidomain structures over which ordinary ARPES measurements perform averages. For instance, in the case of crystal twinning, the complication is caused by the "breathing" kagomé motif of Fe$_3$Sn$_2$ consisting of two different corner-triangle sizes denoted by triangles with different colors in figure 1a, resulting in a three-fold rotation symmetry, rather than six-fold, in the absence of magnetic order [4]. The twinning will be clearly observed when the isoenergy surfaces are away from the high symmetry $k_z = \left(0, \frac{\pi}{c}\right)$ planes as shown by the results of the DFT calculations in figure 1b (U = 1.3 eV); U = 1.3 eV is chosen (figure 1c) as it can reproduce the electron pockets found by ARPES near the Γ point, which will be discussed in detail later. However, the ARPES results reported thus far are predominantly six-fold symmetric [4,18-20], which can only be attained at exactly the high symmetry $k_z = \left(0, \frac{\pi}{c}\right)$ planes for a single domain phase. Tanaka et al. [20] interpreted their ARPES results as being three-fold symmetric for $k_z \neq \left(0, \frac{\pi}{c}\right)$, but the data quality is not sufficient to reach that conclusion unambiguously.

Here, we address the question of discriminating domains with different crystallographic twins, surface termination, and magnetic orientation, by using three forms of photoemission microscopy to characterize $Fe_3Sn_2$, namely, spatially-resolved X-ray photoelectron spectroscopy (XPS), X-ray absorption spectroscopy (XAS), and X-ray magnetic circular dichroism (XMCD), which are used to distinguish physical domains related to surface termination and magnetism, respectively, and show that in fact they are uncorrelated. Separately, by using $\mu$-focused laser ARPES, we probe at very high energy and momentum resolution the band structure close to the Brillouin zone center Γ. In addition to discovering electron pockets, we also observe directly that there are different electronic domains in $Fe_3Sn_2$ rotated around the c-axis by $\delta = m \times 60°, m = \pm(1,3,5)$. By varying the temperature and by comparing the evolving magnetic domain pattern, we conclude that the observed differences in band structure are solely due to twinning and not due to different magnetic states. Direct comparison of the laser $\mu$-ARPES data with synchrotron ARPES data that averages over much larger areas (presented in the SI) shows that the data from synchrotron ARPES corresponds to the average over crystallographic twins and therefore cannot be interpreted without help from either calculations or a spatially resolved technique like $\mu$-ARPES.

**Spatially-resolved XMCD, XAS, and XPS**

We first consider the magnetic domain configuration and different surface terminations, which typical ARPES averages over, by using x-ray photoemission electron microscopy (XPEEM) to obtain spatially resolved XAS and XPS maps of the sample. In our case, the XAS is monitored using photoelectrons with kinetic energy below ~ 1 - 2 eV (low energy secondary electron, 'bulk' sensitive to a depth of a few nm from the surface) while the XPS photoelectrons have a kinetic energy of 96 eV (surface sensitive); the sample was cleaved in vacuum and investigated at 80 K, the base temperature of the cryostat. Figure 2a shows the XMCD map, which is obtained from the differences of XAS collected for left circular (CL) and right circular (CR) polarized photons at the Fe $L_3$-edge, while figure 2b shows the map of the photoelectron yield summed over both polarizations (CL+CR). Figure 2a displays the magnetic domains of the system while figure b confirms that the iron in the bulk of the sample is homogeneous.

Moving to the topic of surface termination, figure 2c shows the XPS map for the same region as in figure 2a and b, obtained for the Sn 3d levels and indicating variations in the Sn intensity. Similar results are observed for a different cleaved surface, shown in figure SI2. Figure 2d displays the intensity histogram from the red square in figure 2c: it is bimodal (in this case well characterized as the sum of two Gaussians), which indicates that there are two possible Sn populations for the surfaces formed upon cleaving $Fe_3Sn_2$. The ratio between the mean intensities for the two Gaussians is $I_1/I_2 = 1.8 \pm 0.6$, where $I_1$ is the brighter intensity and $I_2$ is the darker intensity mean value (follow the derivation in SI). The quantization of the Sn intensity suggested by the bimodal histogram is also apparent from the line scan, also shown in figure 2d, revealing a sharp step in intensity. These phenomena follow because the material consists of (kagomé) Fe bilayers alternating with stanene monolayers. In the first two top-most layers one expects either one or two Fe layers (no stanene layer). Thus, the clear contrast in Sn distribution (figure 2c) arises from stanene-termination for the brighter areas and kagomé-termination for the smaller lower intensity islands. Furthermore, given the much bigger area associated with the larger Sn photoemission intensity, we conclude that the majority of the sample is stanene-terminated, agreeing with the suggestion made in reference [4]. Comparison of figures 2c and 2a also demonstrates that the magnetic domain pattern is uncorrelated with the surface termination. More discussions about the XPEEM and its interpretation can be found in the SI.

The XMCD map reveals a magnetic domain pattern with a hierarchy of characteristic features ranging from 50 to < 1 µm, indicating the need for $\mu$-ARPES to avoid averaging of momentum and energy-resolved states associated with different magnetization directions at 80 K. At the same time, the different surface terminations as established by XPS are characterized by length scales in the same range, implying that $\mu$-ARPES could also detect differences in surface states should they matter for the photon energy chosen.

**Laser $\mu$-ARPES**

We performed laser $\mu$-ARPES (detailed setup described in SI) to probe the electronic structure at high energy and spatial resolution to determine inhomogeneities in the electronic structure linked to magnetic, chemical, and crystallographic domains. The 6.01 eV photon energy limits the Brillouin zone probed to near the Γ point, which nonetheless is of great interest because it enables detailed imaging of the electron pocket seen previously at low resolution with a large beam [4], and predicted by DFT [4,19,20] but not by other band theories [18]. Figure 3a shows our experimental data at 6 K, which reveal this electron pocket with unprecedented clarity for two laser beam positions on the sample.

The Fermi surfaces are located on three rings in $(k_x, k_y)$, with the inner ring clearly three-fold modulated, and the second and third less obviously so, but nonetheless with their most intense feature 60 degrees out-of-phase with the inner ring. The data presented in figure 3a are superpositions of data from both linear laser polarizations. Comparison of the data in figure 3a(i) and (ii) immediately indicates a relative rotation of 180°, which is equivalent to 60° for this rhombohedral material. We thus come to the important finding that the electronic structure of $Fe_3Sn_2$ at 6 K shows two distinct yet equivalent areas (rotated from each other). The positions of these rings are located at the center of the BZ as shown in figure 3b(i). The dispersion curves in figure 3a(iii) and (iv) show the intercepts of three bands with the three Fermi surfaces, and confirms the same 180° rotation in $(k_x, k_y)$ which relates to their respective Fermi surfaces measured in figure 3a(i) and (ii). Two of the bands cross the inner and outermost Fermi surfaces and are labeled as $\alpha$ and $\gamma$ in figure 3a(iii) and (iv), respectively, and merge to form a parabola centered at $\bar{\Gamma}$ with 0.1 eV minimum binding energy. There is also a very sharp feature between $\alpha$ and $\gamma$ labelled $\beta$, which abruptly loses intensity through $\alpha$ around $E_B = 0.025$ eV, as can be seen from the energy-momentum cut image (figure 3a(iii) and (iv)). We also notice a very weak broad band lying at higher binding energy labeled as $\delta$, which has asymmetric intensity similar to that of the $\gamma$ and $\beta$ bands, indicating that it is three-fold symmetric as well.

The two distinct areas rotated by 180° with respect to each other, *i.e.*, twinned domains, can be mapped in real-space by tracing the intensity differences between the left versus right side of the $\alpha$ band in the ARPES cut [figure 3a(iii) and (iv), where data for LH and LV polarizations are summed]. Figure 3c is the resulting domain image, where for $\alpha$ band intensity larger on the left side, we label the sample position in red, otherwise we color it green. In the gradation from green to red, white indicates no intensity difference. In the upper panel of figure 3c, we see that there is a relatively abrupt transition from the "red" to "green" regions, indicating that the twinned domain is rotated discontinuously. A rough measurement of the domain wall angle gives a value of 60°, in agreement with the twin rotation for the crystal structure.

The bulk electronic properties of Fe$_3$Sn$_2$ undergo pronounced evolution on cooling below 80 K, where the spin reorientation to the low temperature phase is complete[5]. Particularly dramatic are the modulation of the carrier density depending on the magnetization direction below 80 K [24] and the temperature dependence of the anisotropic magnetoresistance and planar Hall effect [23], which features a threefold (not sixfold) antisymmetric term as a function of in-plane field direction. To determine whether any of the ARPES features which we see at low temperature could be related to these transport phenomena, we have collected µ-ARPES data at 80 K, which is at the boundary of these characteristic temperatures but still in the regime where the magnetism is dominated by domains with moments along the kagomé planes[6], although with smaller domains compared to 6 K. The lower panel of Figure 3c shows the intensity asymmetry image for 80 K, which is not appreciably different from the 6 K image above it. This is not surprising since based on magnetometry [6], magnetoresistance [5], and magnetic domain imaging[6], the magnetization is in the kagome plane for temperatures below 80 K. The major effect of warming in this case is a broadening of the spectra, and in particular the loss of the very sharp $\beta$ bands as seen in figure 4a, which we discuss in greater detail below. In the SI, we show that the earlier synchrotron ARPES data are consistent with an average over the twin domain data obtained from laser µ-ARPES. We attribute the three-fold pattern itself to originate from probing off high-symmetry $k_z$ plane, yet still close to $k_z = 0$ plane (see SI for further explanation), confirming the bulk nature of the spectral data; high symmetry probing at $k_z = 0$ plane or $k_z = \pi$ plane will generate a six-fold pattern regardless of the "breathing" pattern of the kagomé.

**Comparison to DFT calculations**

We now take advantage of the high real and reciprocal space resolution of our single domain ARPES to investigate the effect of correlations on the electronic band structure of Fe$_3$Sn$_2$. The first question concerns the effective on-site Coulomb potential $U$, which should be included in the DFT + $U$ band structure calculations. Electron pockets centered on Γ and close to the Fermi level exist for a large range of $U$ values ($U = 0.0 - 3.0$ eV) (see figure SI7). However, the distance $\Delta E$ to the Fermi energy from the bottom of the bands for the pocket varies strongly with $U$, as do the Fermi vectors $k_F$, which can be seen in figure 1c and SI7. The best description of the $\alpha$ and $\gamma$ bands, taking into account both the band bottom and $k_F$, occurs for $U = 1.3$ eV, differing from the values $U = 0.5$ eV [4] or $U > 2$ eV [25] selected for previous DFT calculations.

We display the DFT cut along $\overline{M}\overline{\Gamma}\overline{M'}$, for $k_z = 0 - 0.05\,\pi$, $U = 1.3$ eV in figure 3b(ii). The cut is dominated by parabolic bands coexisting with a relatively flat band located around 15 meV above $E_F$, and supports the assertion that $\alpha$ and $\gamma$ originate from bulk states smeared due to $k_z$ broadening [26], and are asymmetric (3-fold instead of 6-fold) due to slight off $k_z = 0$ probing. We have labelled the features in the cut accordingly: the $k_z$ broadening yields two visible branches cutting the Fermi level along the $\overline{M}\overline{\Gamma}$ trajectory, while there is a broader, merged $\alpha$ – like dispersion along $\overline{\Gamma}\overline{M'}$. The Fermi velocities do not precisely match those obtained by ARPES but can be tuned via small adjustments to the chemical potential on account of the apparent hybridization with the flat band very close to the Fermi level. We note that irrespective of the choice of U significant differences remain between ARPES and DFT. None of the existing DFT calculations reproduces the very small gap between the bottom of the α and γ bands and the top of δ; additionally, the sharp β band found by ARPES near the Fermi level is not seen in calculations.

**Sharp but marginal quasiparticles at low temperatures**

Key parameters of quasiparticles which include their Fermi surface areas, mass, and scattering rates, are calculated from our experimental data. We display the representative dispersion cuts for the individual linear photon polarizations in figure 4a(i) and (iii) for horizontal (LH) polarization at 6 and 70 K, respectively, and in figure 4a(ii) and (iv) for vertical (LV) polarization at 6 and 70 K, respectively. LH displays mostly the $\alpha$ band, indicating it is an even orbital with respect to the incident plane of the experiment, while the $\beta, \gamma$, and $\delta$ bands are odd parity orbitals. In Table I we summarize for the three pockets the effective mass at the Fermi level calculated as $m^* = \hbar k_F (v_F)^{-1}$, where the Fermi velocity is $v_F = \frac{1}{\hbar}\frac{\partial E}{\partial k}\Big|_F$, the area of the electron pocket converted into the de Haas-van Alphen (dHvA) frequency as $f(T) = \frac{\hbar}{2\pi e} \times area$, and the scattering lengths of the quasiparticle defined as $\lambda = \frac{1}{\Delta k}$, where $\Delta k$ is the half width of the momentum distribution curve (MDC) peaks (at 6 K and at the Fermi level). The area of the electron pockets is calculated by tracing the visible peaks of the corresponding band at the Fermi level and assuming a circular extrapolation to make closed contours. More details on the EDC of the bands from $k_F$ to band minimum can be found in figure SI5.

For the $\beta$ band, $\lambda = 155 \pm 5$ Å at 6 K, a remarkably long mean free path for a topological (bulk) metal and comparable to that for the much simpler surface Dirac quasiparticles for the *n*-type Bi$_2$Te$_3$ topological insulator (~190 Å) [27]; for $\alpha$, $\lambda$ is reduced by a factor of two to ~ 85 Å but is still long by the same standards. These values are consistent with the residual ($T = 2$ K) resistivity $\rho_{xx}(0) = 4.35\ \mu\Omega.cm$ and a carrier density of $1.8 \times 10^{28}\ m^{-3}$ from transport data [5,24] for samples similar to ours, grown by the same method by the same group. In particular, applying the Drude formula $\sigma_0 = \frac{ne^2\tau}{m}$ for conductivity, we obtain a scattering time $\tau \sim 4.5 \times 10^{-14}$ s by assuming that the carriers have an effective mass $m = m_e$. For an averaged Fermi velocity of ~1.2 eV- Å, $\tau$ converts to a mean free path $l = v_F\tau \sim 82$ Å, which is of the order of the ARPES results for the $\alpha$ and $\beta$ bands.

It is also interesting to ask whether there is bulk evidence for small Fermi surfaces similar to those which we have discovered. The answer is yes: the calculated de Haas van Alphen (dHvA) frequency 254 T of the $\beta$ band, whose Fermi surface is centered on $\Gamma$, is in reasonable agreement with that measured for reference [7] (~200 T) and attributed to the "Dirac ring" emanating from the putative Dirac point at the $\overline{K}$ point and 0.1 eV below the Fermi surface. Given that DFT accounts for electron pockets centered on $\Gamma$ while explaining the Dirac-like features in terms of an overlap of surface and bulk bands[20], a re-interpretation of the observed dHvA frequency is needed.

Moving below $E_F$, the $\beta$ band is significantly broadened and effectively disappears for binding energies in excess of approximately 20 meV. Below this energy, the half-width at half-maxima $\Delta k(E)$ of the peaks in the MDC increases linearly with $|k(E) - k_F|$, with slope $b_\beta = 0.361 \pm 0.062$, as shown in figure 4. For the much steeper $\alpha$ band, there is a similar finding: $\Delta k(E)$ rises linearly with $|k(E) - k_F|$. The fitted slope $b_\alpha = 0.353 \pm 0.071$ is indistinguishable from both that for the $\beta$ band, and is also not far from $b_\pi = \frac{1}{\pi}$, for which the relative quasiparticle wavelength $2\pi/|k(E) - k_F|$ is precisely twice the quasiparticle scattering length $1/\Delta k(E)$. Because of the linearity of the quasiparticle dispersions (upper frame of Fig. 4b) over the same momentum ranges, the low temperature (6 K) results are all consistent with a marginal Fermi liquid hypothesis [28,29], for which the imaginary part of the self-energy is $Im(\Sigma) = \Delta E = \max(|E_B|, k_B T) = \max(b\hbar v_F |k(E) - k_F|, b'k_B T)$, which translates to $\Delta k(E) = \max(b|k(E) - k_F|, b'k_B T/\hbar v_F)$ (1), where $b$ and $b'$ are coupling constants of order unity, where the initial theoretical Ansatz was $\frac{b'}{b} = 1$. Experiments have yielded a variety of values for $b$ in strongly correlated systems. For example, for the cuprate superconductor $Bi_2Sr_2CaCu_2O_{8+\delta}$ (BSCCO), $b_{BSCCO} = 0.75$[29], while for $BaCr_2As_2$, $b_{BaCr_2As_2}$ depends strongly on the band chosen, with resulting values from 0.35 to 0.63 [30], and for the ferropnictides $b$ varies throughout the range between 0 and 1.7 [31,32].

Eq.(1) implies that a temperature increase from $T_{base} = 6$ to 70 K, as in figure 4a should cause, for small $E$ and $\frac{b'}{b} = 1$, $\Delta k(E)$ to increase by $b'k_B(T - T_{base})/\hbar v_F = 1.3 \times 10^{-3}$ Å$^{-1}$ and $2.7 \times 10^{-3}$ Å$^{-1}$ for the $\alpha$ and $\beta$ bands respectively. This is much more modest than what we actually observe for $Fe_3Sn_2$, where at $E_F$, $\Delta k$ for $\alpha$ grows by $(8 \pm 1) \times 10^{-3}$ Å$^{-1}$, which is $b'/b = 6$ times larger than the theoretically expected ratio of unity and 2.4 times bigger than $b'/b$ observed for BSCCO [31,32], while corresponding to a very large scattering rate, $\Delta E = \hbar \Delta k * v_F \approx 20$ meV $\approx 3.3$ $k_B$T. At the same time, any distinction between the $\beta$ and $\gamma$ bands, which near the Fermi level are simply displaced by 20 meV with respect to each other, vanishes, which is consistent with the $\Delta E$ measured for the $\alpha$ band at 70 K, and also the fact that at 6 K, the $\beta$ band becomes invisible for binding energies above a similar energy.

The conclusion is that $E/T$ scaling is roughly obeyed in the sense that marginal quasiparticles for both the $\alpha$ and the $\beta$ bands are distinguishable or not near the Fermi surface to an extent consistent with Eq(1). However, the value for $b'/b$ is unusually large, and the energy-momentum diagram is very peculiar in that the much broader $\gamma$ band seems much more associated with the minimum near $\Gamma$ than the short, 20 meV stretch of $\beta$ band visible near the Fermi surface, which simply looks like a sharp copy of the $\gamma$ band displaced upwards by the same ~20 meV towards $E_F$. These observations, not easily accounted for by DFT, together suggest a many-body origin for the $\beta$ band. A possible scenario is that there is strong hybridization between flat and $\gamma$ band states near the bare Fermi surface which results in interaction-induced (Mott-like) localization that reduces the free carrier density; $k_F$ for the electron pockets seen here would then be correspondingly reduced by $\delta k_F$, and the resulting ($\beta$) band will be raised towards $E_F$ by $\Delta = \hbar v_F \delta k_F$. There will of course still be broad spectral weight from the unrenormalized ($\gamma$) bands at the original $k_F$, and for binding energies in excess of $\Delta$, the localized electrons rejoin the Fermi sea and the spectral weight will reside primarily near the $\gamma$ band as might be calculated using DFT. Warming would similarly act to delocalize the carriers, and thus cause the loss of the $\beta$ "band".

| Band | Fermi surface area ($\text{Å}^{-2}$) | dHvA frequency (T) | Fermi velocity $v_F$ ($\text{Å} \cdot s^{-1}/\hbar$) | Effective mass ($m^*/m_e$) | Scatt. length $\lambda$ (Å) |
|---|---|---|---|---|---|
| $\alpha$ | $0.015 \pm 0.005$ | $157 \pm 20$ | $1.62 \pm 0.05$ | $0.36 \pm 0.06$ | $85 \pm 5$ |
| $\beta$ | $0.024 \pm 0.005$ | $254 \pm 20$ | $0.78 \pm 0.01$ | $1.01 \pm 0.08$ | $155 \pm 20$ |
| $\gamma$ | $0.041 \pm 0.002$ | $433 \pm 20$ | $0.79 \pm 0.04$ | $1.10 \pm 0.10$ | $38 \pm 5$ |

Table I Summary of the calculated quantities (at 6 K): Fermi surface area, dHvA frequency, Fermi velocity, effective mass, and scattering length, calculated at the Fermi surface of the corresponding $\alpha$, $\beta$, and $\gamma$ bands.

**Conclusions**

We have discovered electron pockets labeled as $\alpha$, $\beta$, and $\gamma$ centered at the $\Gamma$ point in Fe$_3$Sn$_2$ by employing state-of-the-art $\mu$-focused laser ARPES to overcome the averaging effects of crystallographic twinning in Fe$_3$Sn$_2$ crystals. These electron pockets are not found in the simple tight-binding models proposed for Fe$_3$Sn$_2$, which also posit doubled Dirac cones near the K point, while they can be reproduced by DFT calculations, where no Dirac cones are predicted (see SI for detailed discussion). The $\alpha$ and $\gamma$ electron pockets are very sensitive to the Coulomb interaction parameter $U$ in DFT+U calculations, and are well reproduced by $U = 1.3$ eV. The same calculations also yield a flat band about 20 meV above the Fermi level for the pockets.

Our combination of high-resolution spectroscopy with single domain selection provides an unprecedented opportunity to examine the quasiparticles in a strongly correlated kagome metal. This means that we are able to measure scattering lengths as well as Fermi surface areas. The former is in semiquantitative agreement with the mean free path obtainable from residual resistance data, while the Fermi surface areas are near de Haas van Alphen results previously interpreted to derive from the Dirac cones not found by DFT.

At low temperature, once the quasiparticle Green's functions are examined away from the regime where their widths are dominated by elastic scattering, it appears that their widths in momentum space scale with their momenta, implying marginal rather than ordinary Fermi liquid behaviour, and associated strong correlation physics. However, the abrupt disappearance of the sharpest band at the very modest binding energy of 20 meV, and quasiparticle broadening greatly in excess of $k_B T$ on warming to 70 K indicate that the kagome ferromagnet $Fe_3Sn_2$ is hosting more than "conventional" strong correlations. The flat band seen in DFT just above the Fermi energy is a distinguishing feature: virtual and thermal occupancy of this band could greatly enhance electron-electron scattering that we see directly in our data, and also account for formation of the peculiarly sharp and displaced $\beta$ band which appears below 70 K together with numerous anomalous magnetotransport properties. We close by noting that there is an enormous literature dealing with the Anderson and Kondo lattices where dispersive bands coexist with bands which are flat on account of small underlying atomic orbitals[33]. Our $\beta$ band results, which we can only explain at a qualitative level, suggest further theoretical and experimental studies to examine interaction effects where dispersive bands cross bands which are flat for reasons of lattice geometry, as exhibited for example by kagome systems, rather than small orbital extent [34].

**Acknowledgement**

S.A.E, D.G.-M., O.V.Y., M.S. acknowledge the support from NCCR-MARVEL funded by Swiss National Science Foundation, S.A.E. acknowledges the European Union's Horizon 2020 research and innovation programme under the Marie Skłodowska-Curie grant agreement No 701647, G.A. acknowledges the European Research Council HERO Synergy grant SYG-18 810451. The laser-ARPES work at the University of Geneva was supported by the SNSF grants 2000020_165791, 200020_184998. Part of this work was performed at the Surface/Interface: Microscopy (SIM) beam line of the Swiss Light Source, Paul Scherrer Institut, Villigen, Switzerland. All first-principles calculations were performed at the Swiss National Supercomputing Centre (CSCS) under the projects s1008 and mr27. We acknowledge Neeraj Kumar for taking the XMCD-PEEM images, and Markus Müller for helpful discussions.

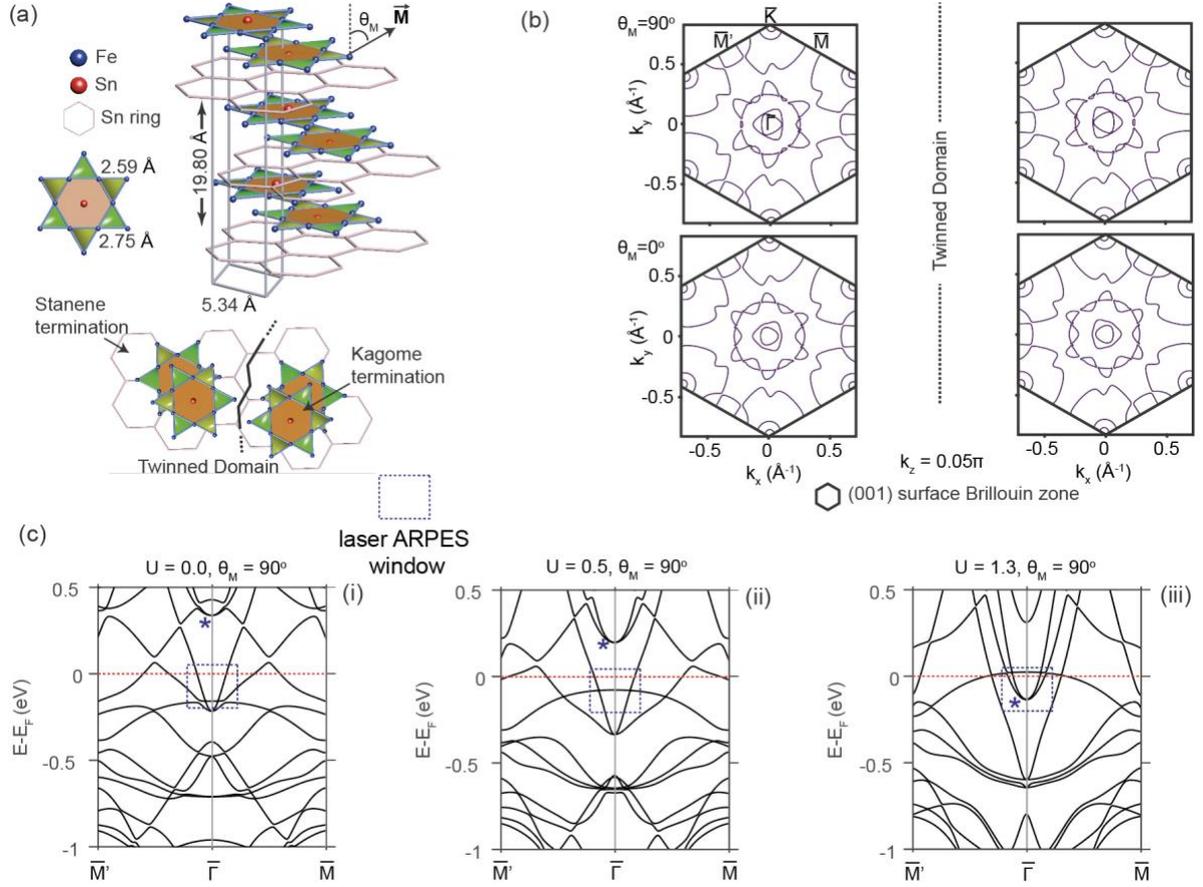

**Figure 1.** a. Crystal structure of Fe$_3$Sn$_2$ displaying a stanene layer sandwiched between kagomé bilayers. $\theta_M$ denotes the angle between the magnetic moment $\vec{M}$ and the *c*-axis. Twin domains may occur when the system is rotated around the *c*-axis. b. Calculated Fermi surface at $k_z = 0.05\pi/c$ for $U = 1.3$ eV showing a three-fold pattern (left) and its twinned image (right) when the magnetic moment is pointing off the *c*-axis ($\theta_M = 90°$) and along the c-axis ($\theta_M = 0°$). c. Results of DFT calculations for $k_z = 0$ with representative $U = 0$, 0.5, and 1.3 eV values showing the presence of an electron pocket (denoted by asterisks), which is investigated by μ-ARPES (marked by the area within the dashed-line).

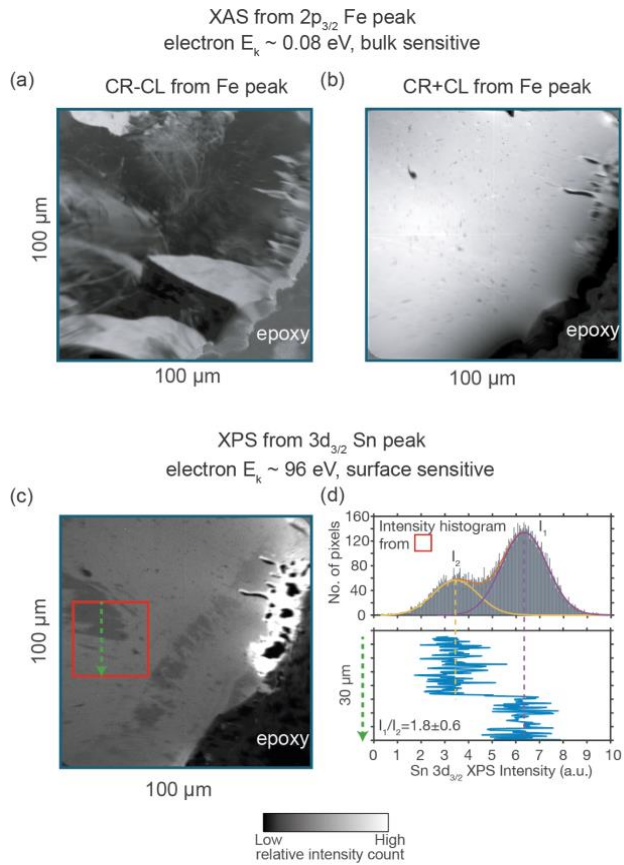

**Figure 2**. a. XPEEM magnetic contrast (CR-CL intensity) map taken at the Fe $L_3$ peak, whose signal is 'bulk'-sensitive, showing a complex magnetic domain configuration in an area where b. panel (CR+CL intensity) shows a homogeneous contrast, indicating a homogeneous phase of the sample. c. XPEEM XPS map of the sample at the Sn $3d_{3/2}$ peak, whose signal is surface sensitive, showing a domain structure distinct from the magnetic domain shown in panel a., which is attributed to different cleaving terminations, *i.e.*, two different surface terminations. This shows that the magnetic domain has no correlation to the surface termination. Panel d. shows that the intensity difference found in panel c. is step-like with intensity ratio of $1.8 \pm 0.6$.

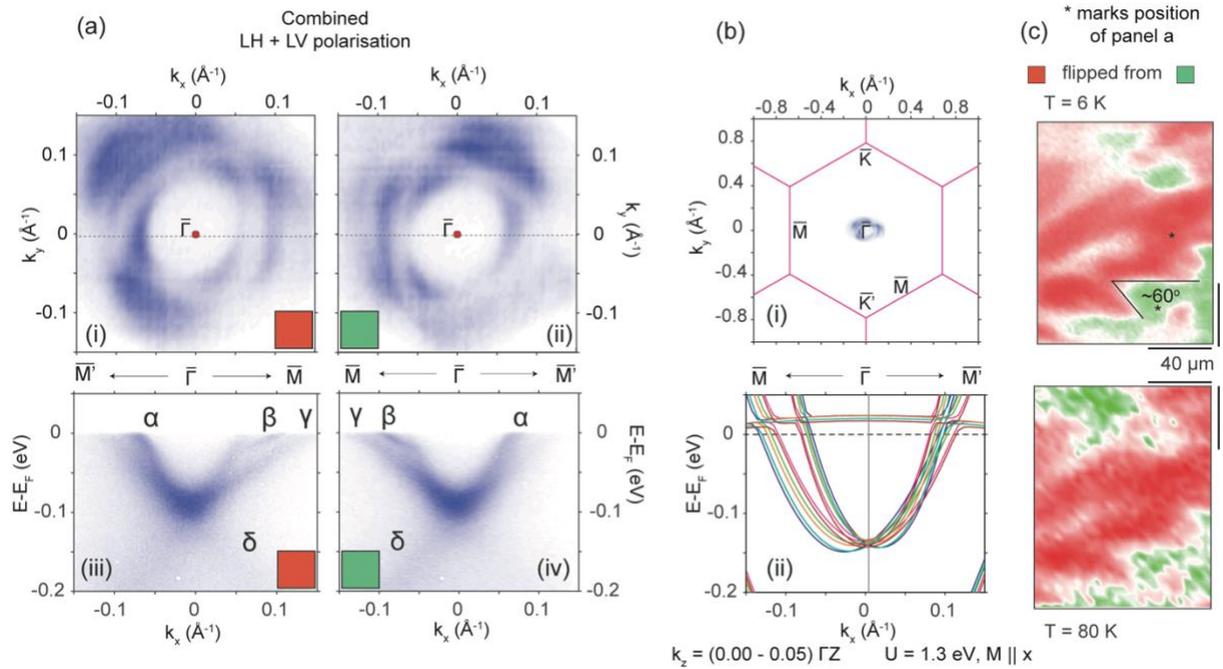

**Figure 3.** a. Laser ARPES data for Sn-terminated Fe₃Sn₂. (i)-(ii) Fermi surfaces showing a three-fold pattern with an electron pocket at the $\bar{\Gamma}$ point and three petals, which are rotated relative to each other distinguishing the crystal twin domains. (iii)-(iv) Energy versus momentum cut of one of the petals (dashed line at (i) and (ii), $\overline{M\Gamma M'}$ direction) showing the electron pocket band labeled $\alpha$, a band labeled $\beta$, which abruptly disappears through $\alpha$, a petal labeled $\gamma$ sharing a bottom with the $\alpha$ band, and lastly a broad and weak band referred to as $\delta$. b (i) Illustration of the reciprocal space probed with laser ARPES, which spans a small portion of the total BZ. b (ii) DFT calculation along $\overline{M\Gamma M'}$ direction for $k_z = 0 - 0.05\,\pi$ as illustration of $k_z$ broadening and asymmetrization, agreeing tentatively on the formation of $\alpha$ and $\gamma$ bands. c. Twin domains showing a relatively unchanged shape at 6 K as compared to 80 K. The magnetisation points along the *ab*-plane at both temperatures but the magnetic domain size is known to be different at these two temperatures, with large domains at temperatures below 64 K and with relatively smaller domains at 80 K [6].

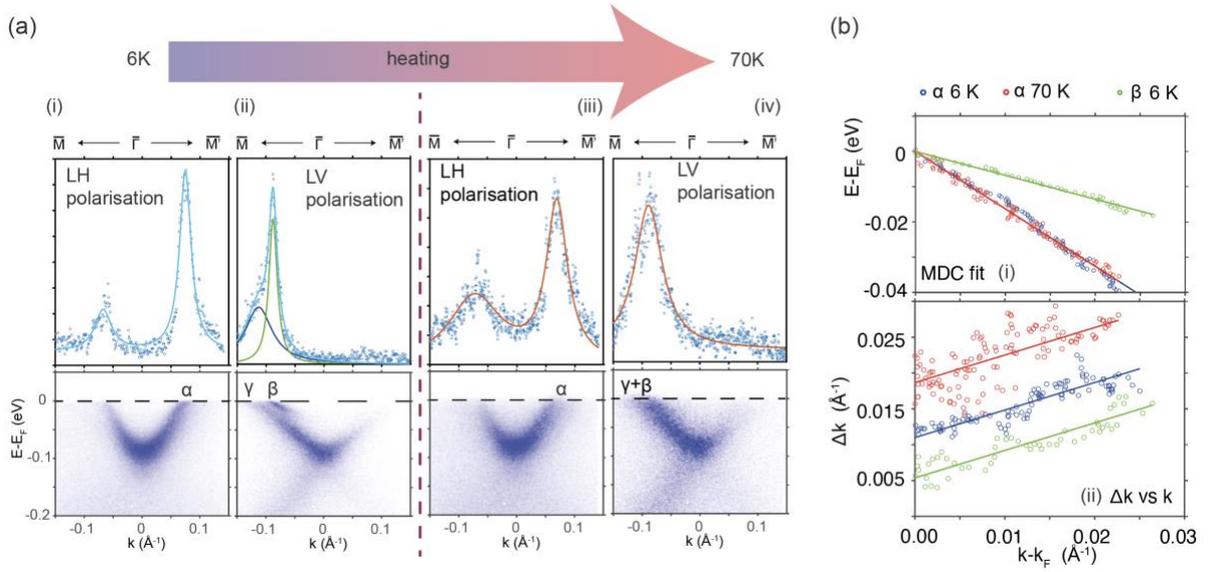

**Figure 4.** (a) $\overline{M}\overline{\Gamma}\overline{M}'$ cut from LH polarization displays mostly band labeled $\alpha$ at 6 K (i) and 70 K (iii), with MDC at the Fermi energy and the fitting result. $\overline{M}\overline{\Gamma}\overline{M}'$ cut from LV polarization reveals mostly the band labeled $\beta$ and $\gamma$ at 6 K (ii) and 70 K (iv), with MDC at the Fermi energy and the fitting result. (b) (i) Plot of the fitted peak position from MDC for $\alpha$ (blue 6 K and red 70 K) and $\beta$ (green 6 K) with resulting linear fit super-imposed. (ii) Plot of peak width $\Delta k$ vs $k - k_F$ showing a same-gradient linear behaviour close to the Fermi level for $\alpha$ (blue 6 K and red 70 K) and $\beta$ (green 6 K) indicating a marginal Fermi liquid. Upon heating to 70 K we can see that the line is shifted as described by Eq. (1) in text.

**Supporting Information**

### A. Sample growth and structural characterization

$Fe_3Sn_2$ crystalizes in a rhombohedral structure with space group $R\bar{3}m$, with crystal axis $a, b = 5.34$ Å, $c = 19.80$ Å, and $\gamma = 120°$. $Fe_3Sn_2$ single crystals were grown by a vapor transport method. Stoichiometric iron powder (Alfa Aesar, > 99.9%) and tin powder (Alfa Aesar, > 99.9%) were placed into an evacuated quartz tube. Then the tube was annealed at 800°C for 7 days before quenching in icy water. The crystal structure of the polycrystalline $Fe_3Sn_2$ precursor is confirmed by the x-ray diffraction (XRD) data shown in Fig. SI1a. The obtained polycrystalline $Fe_3Sn_2$ precursor was thoroughly ground and sealed with $I_2$ (~ 4 mg/cm$^3$) in a quartz tube 1 cm in diameter and 16 cm in length. $Fe_3Sn_2$ single crystals were obtained under a temperature gradient of 650 °C (source) to 720 °C (sink) for two weeks.

The cleaved surface for laser ARPES is confirmed to be (001), the *ab* plane, from low energy electron diffraction (LEED) measurements on a freshly cleaved surface, as shown in Fig. SI1b. The typical cleaved surface is displayed in figure SI1.c indicating a flat surface suitable for ARPES measurements.

### B. Magnetic domain and surface termination probed by X-ray photoemission electron microscopy (XPEEM)

The surface termination of the $Fe_3Sn_2$ samples is probed by mapping the Sn $3d_{3/2}$ and Fe $2p_{3/2}$ local photoemitted electron intensities using the spectroscopic capabilities of the SPELEEM III instrument (Elmitec GmbH) at the Surface/Interface Microscopy (SIM) beamline of the Swiss Light Source. The photon excitation energy was varied such that the kinetic energy of the detected electrons was fixed at 96.8 eV in order to achieve a high surface sensitivity (~2 Å according to the universal curve [1]). The photoemission peaks are determined by recording XPEEM image sequences with x rays, linearly polarized in the plane of the sample, with varying photon energy. The samples are cleaved *in situ* prior to the XPEEM measurements at 80 K, base temperature of the cryostat. A typical result is shown in Fig. SI2a, displaying a clear cleaved surface and an area with glue residue and a crack. The XPS intensity distributions for the Sn $3d_{3/2}$ and Fe $2p_{3/2}$ peaks are shown in Fig. SI2b, c, respectively. Electrons emitted from the Sn $3d_{3/2}$ orbitals have two distinct intensities on the spatially resolved map in Fig. SI2b, revealing a large region of higher intensity together with smaller islands of lower intensity. The iron emission map shown in Fig. SI2c is more homogeneous and has a much lower signal to noise ratio, expected from the lower non-resonant x-ray absorption of Sn, Z = 50, when compared to Fe, Z = 26, which prevents the observation of the smaller islands as in panel (b). The XPS spectra displayed in panel (c) are obtained by integrating over an area larger than that for the Sn XPS peak shown in panel (b). We assign the bright area in Fig. SI2b as stanine-terminated and the darker area as kagomé-terminated. However, we cannot determine the exact kagomé termination, *i.e.*, whether it is one or two Fe layers (Fig. SI2d.i or ii) within the intensity resolution of our measurements. The cleaved surface discussed here is different from the photoemission experiment presented in figure 2 of the main manuscript. A comparison between these two sets of data shows that the area attributed to kagomé-terminated can vary in size and is not limited to only small areas as shown in Fig. 2b.

The magnetic domain configuration was determined using the x-ray magnetic circular dichroic (XMCD) effect. XPEEM images were recorded by exciting the sample with circularly polarized x-rays with the photon energy tuned to the Fe $L_3$ edge and measuring the intensity of the low energy secondary electrons (with a probing depth of about 3-5 nm). Magnetic contrast maps were finally obtained by pixelwise division of two XPEEM images recorded with opposite x-ray helicities.

**Intensity ratio of XPS signal**

The PEEM micrographs showing different terminations in Fe3Sn2 can be numerically analysed as following:
1. We assume that the photon penetration depth is much longer than the escape depth of the most energetic photo-electron, *i.e.*, an electron at the Fermi level. Thus, we ignore the exponential decay component of the photon intensity.
2. Assumption 1 leads to an expression of photo-electron intensity from a given atom called $a$, on layer called $b$, with distance $d$ from the surface as

$$I_{ab} = C_{ab} n_{ab} \exp\left(-\frac{d}{\lambda}\right)$$

where $\lambda$ is the escape depth of the electron, $C_{ab}$ is the proportionality constant that contains the photon cross section of atom $a$ at layer of type $b$, and $n_{ab}$ is the atom $a$ density at layer of type $b$.
3. Generalizing (2), we can express the total intensity from atoms *a* from all layers of type *b* as a summation

$$I_{ab\;total} = C_{ab} n_{ab} \sum_{i=1}^{\infty} \exp\left(-\frac{d_i}{\lambda}\right)$$

where we assume the proportionality constant $C_{ab}$ to be the same for all similar layers at different depths.

With these three points we can start to write down the expression of XPS intensity from both Sn and Fe in total. First, we notice that there are three possible terminations and call them double kagome termination ($S_{kk}$), single kagome termination ($S_k$), and stanene termination ($S_s$) as shown in figure SI2d.

**Double kagome termination ($S_{kk}$) case**
The total Sn and Fe intensity from this termination can be expressed as

$$
\begin{aligned}
I_{\text{Sn},S_{kk}} &= \sum_{m=0}^{\infty} \left(\exp\left(-m\frac{2z_{ks}+z_{kk}}{\lambda}\right) \exp\left(-\frac{z_{ks}+z_{kk}}{\lambda}\right) C_{\text{Sn,s}} n_{\text{Sn,s}}\right) \\
&+ \sum_{m=0}^{\infty} \left(\exp\left(-m\frac{2z_{ks}+z_{kk}}{\lambda}\right) C_{\text{Sn,k}} n_{\text{Sn,k}}\right) \\
&+ \sum_{m=0}^{\infty} \left(\exp\left(-m\frac{2z_{ks}+z_{kk}}{\lambda}\right) \exp\left(-\frac{z_{kk}}{\lambda}\right) C_{\text{Sn,k}} n_{\text{Sn,k}}\right) \\
&= \frac{\exp\left(-\frac{z_{ks}+z_{kk}}{\lambda}\right) C_{\text{Sn,s}} n_{\text{Sn,s}} + C_{\text{Sn,k}} n_{\text{Sn,k}} + \exp\left(-\frac{z_{kk}}{\lambda}\right) C_{\text{Sn,k}} n_{\text{Sn,k}}}{1 - \exp\left(-\frac{2z_{ks}+z_{kk}}{\lambda}\right)}
\end{aligned}
$$

$$I_{\text{Fe},S_{kk}} = \sum_{m=0}^{\infty} \left( \exp\left(-m\frac{2h_{ks} + h_{kk}}{\lambda}\right) C_{\text{Fe,k}} n_{\text{Fe,k}} \right)$$

$$+ \sum_{m=0}^{\infty} \left( \exp\left(-m\frac{2h_{ks} + h_{kk}}{\lambda}\right) \exp\left(-\frac{h_{kk}}{\lambda}\right) C_{\text{Fe,k}} n_{\text{Fe,k}} \right)$$

$$= \frac{C_{\text{Fe,k}} n_{\text{Fe,k}} + \exp\left(-\frac{h_{kk}}{\lambda}\right) C_{\text{Fe,k}} n_{\text{Fe,k}}}{1 - \exp\left(-\frac{2h_{ks} + h_{kk}}{\lambda}\right)}$$

**Single kagome termination ($S_k$) case**

$$I_{\text{Sn},S_k} = \sum_{m=0}^{\infty} \left( \exp\left(-m\frac{2z_{ks} + z_{kk}}{\lambda}\right) \exp\left(-\frac{z_{ks}}{\lambda}\right) C_{\text{Sn},s} n_{\text{Sn},s} \right)$$

$$+ \sum_{m=0}^{\infty} \left( \exp\left(-m\frac{2z_{ks} + z_{kk}}{\lambda}\right) \exp\left(-\frac{2z_{ks}}{\lambda}\right) C_{\text{Sn},k} n_{\text{Sn},k} \right)$$

$$+ \sum_{m=0}^{\infty} \left( \exp\left(-m\frac{2z_{ks} + z_{kk}}{\lambda}\right) C_{\text{Sn},k} n_{\text{Sn},k} \right)$$

$$= \frac{\exp\left(-\frac{z_{ks}}{\lambda}\right) C_{\text{Sn},s} n_{\text{Sn},s} + \exp\left(-\frac{2z_{ks}}{\lambda}\right) C_{\text{Sn},k} n_{\text{Sn},k} + C_{\text{Sn},k} n_{\text{Sn},k}}{1 - \exp\left(-\frac{2z_{ks} + z_{kk}}{\lambda}\right)}$$

$$I_{\text{Fe},S_k} = \sum_{m=0}^{\infty} \left( \exp\left(-m\frac{2h_{ks} + h_{kk}}{\lambda}\right) \exp\left(-\frac{2h_{ks}}{\lambda}\right) C_{\text{Fe},k} n_{\text{Fe},k} \right)$$

$$+ \sum_{m=0}^{\infty} \left( \exp\left(-m\frac{2h_{ks} + h_{kk}}{\lambda}\right) C_{\text{Fe},k} n_{\text{Fe},k} \right)$$

$$= \frac{\exp\left(-\frac{2h_{ks}}{\lambda}\right) C_{\text{Fe},k} n_{\text{Fe},k} + C_{\text{Fe},k} n_{\text{Fe},k}}{1 - \exp\left(-\frac{2h_{ks} + h_{kk}}{\lambda}\right)}$$

**Stanene termination ($S_s$) case**

$$I_{\text{Sn},S_s} = \sum_{n=0}^{\infty} \left( \exp\left(-n\frac{2z_{ks} + z_{kk}}{\lambda}\right) C_{\text{Sn},s} n_{\text{Sn},s} \right)$$

$$+ \sum_{n=0}^{\infty} \left( \exp\left(-n\frac{2z_{ks} + z_{kk}}{\lambda}\right) \exp\left(-\frac{z_{ks}}{\lambda}\right) C_{\text{Sn},k} n_{\text{Sn},k} \right)$$

$$+ \sum_{n=0}^{\infty} \left( \exp\left(-n\frac{2z_{ks} + z_{kk}}{\lambda}\right) \exp\left(-\frac{z_{ks} + z_{kk}}{\lambda}\right) C_{\text{Sn},k} n_{\text{Sn},k} \right)$$

$$= \frac{C_{\text{Sn},s} n_{\text{Sn},s} + \exp\left(-\frac{z_{ks}}{\lambda}\right) C_{\text{Sn},k} n_{\text{Sn},k} + \exp\left(-\frac{z_{ks} + z_{kk}}{\lambda}\right) C_{\text{Sn},k} n_{\text{Sn},k}}{1 - \exp\left(-\frac{2z_{ks} + z_{kk}}{\lambda}\right)}$$

$$I_{\text{Fe},S_s} = \sum_{n=0}^{\infty} \left( \exp\left(-n\frac{2h_{ks}+h_{kk}}{\lambda}\right) \exp\left(-\frac{h_{ks}}{\lambda}\right) C_{\text{Fe},k} n_{\text{Fe},k} \right)$$

$$+ \sum_{n=0}^{\infty} \left( \exp\left(-n\frac{2h_{ks}+h_{kk}}{\lambda}\right) \exp\left(-\frac{h_{ks}+h_{kk}}{\lambda}\right) C_{\text{Fe},k} n_{\text{Fe},k} \right)$$

$$= \frac{\exp\left(-\frac{h_{ks}}{\lambda}\right) C_{\text{Fe},k} n_{\text{Fe},k} + \exp\left(-\frac{h_{ks}+h_{kk}}{\lambda}\right) C_{\text{Fe},k} n_{\text{Fe},k}}{1 - \exp\left(-\frac{2h_{ks}+h_{kk}}{\lambda}\right)}$$

where $C_{\text{Sn},s}$ is the proportionality constant for Sn atoms in the stanene layer, $n_{\text{Sn},s}$ is the Sn density in the stanene layers, $C_{\text{Sn},k}$, $C_{\text{Fe},k}$ are the proportionality constants for Sn, Fe atom in the kagome layer, respectively, $n_{\text{Sn},k}$, $n_{\text{Fe},k}$ are the Sn, Fe densities at the kagome layer, respectively, $h_{kk}$ is the distance of Fe layers between two adjacent kagome layers (the kagome bilayer), $h_{ks}$ is the distance of Fe layers in the kagome layer to the nearest stanene layer, $z_{kk}$ is the distance of Sn atom layers between two adjacent kagome layers (the kagome bilayer), $z_{ks}$ is the distance of Sn atom layers from a kagome layer to the nearest stanene layer.

From the expression above, we can conclude that $\boldsymbol{I_{\text{Fe},S_s} < I_{\text{Fe},S_k} < I_{\text{Fe},S_{kk}}}$ based on the following relation

$$\frac{\exp\left(-\frac{h_{ks}}{\lambda}\right) C_{\text{Fe},k} n_{\text{Fe},k} + \exp\left(-\frac{h_{ks}+h_{kk}}{\lambda}\right) C_{\text{Fe},k} n_{\text{Fe},k}}{1 - \exp\left(-\frac{2h_{ks}+h_{kk}}{\lambda}\right)}$$

$$< \frac{\exp\left(-\frac{2h_{ks}}{\lambda}\right) C_{\text{Fe},k} n_{\text{Fe},k} + C_{\text{Fe},k} n_{\text{Fe},k}}{1 - \exp\left(-\frac{2h_{ks}+h_{kk}}{\lambda}\right)} < \frac{C_{\text{Fe},k} n_{\text{Fe},k} + \exp\left(-\frac{h_{kk}}{\lambda}\right) C_{\text{Fe},k} n_{\text{Fe},k}}{1 - \exp\left(-\frac{2h_{ks}+h_{kk}}{\lambda}\right)}$$

According to this model, we should be able to see the Fe peak contrast between different terminations. However, our data display a rather homogenous signal whose origin is already explained in the previous section.

Meanwhile, for Sn intensity, we can posit that
$$n_{\text{Sn},s} \approx 2 n_{\text{Sn},k}$$
on account of the twice higher density of Sn in the stanene rather than in the kagome layer. And finally, it is also reasonable to assume
$$C_{\text{Sn},s} \approx C_{\text{Sn},k}$$
to obtain

**Double kagome termination ($S_{kk}$) case**

$$I_{\text{Sn},S_{kk}} \approx C_{\text{Sn},k} n_{\text{Sn},k} \frac{2\exp\left(-\frac{z_{ks}+z_{kk}}{\lambda}\right) + 1 + \exp\left(-\frac{z_{kk}}{\lambda}\right)}{1 - \exp\left(-\frac{2z_{ks}+z_{kk}}{\lambda}\right)}$$

**Single kagome termination ($S_k$) case**

$$I_{\text{Sn},S_k} \approx C_{\text{Sn},k} n_{\text{Sn},k} \frac{2\exp\left(-\frac{z_{ks}}{\lambda}\right) + \exp\left(-\frac{2z_{ks}}{\lambda}\right) + 1}{1 - \exp\left(-\frac{2z_{ks}+z_{kk}}{\lambda}\right)}$$

**Stanene termination ($S_s$) case**

$$I_{\text{Sn},S_s} \approx C_{\text{Sn,k}} n_{\text{Sn,k}} \frac{2 + \exp\left(-\frac{z_{ks}}{\lambda}\right) + \exp\left(-\frac{z_{ks} + z_{kk}}{\lambda}\right)}{1 - \exp\left(-\frac{2z_{ks} + z_{kk}}{\lambda}\right)}$$

With these assumptions, we have Sn intensity relation for different terminations as

$$I_{\text{Sn},S_{kk}} < I_{\text{Sn},S_k} < I_{\text{Sn},S_s}$$

From the intensity relation of both Sn and Fe, we can conclude that they are inversely related to each other.

We investigate the Sn intensity ratio between terminations to correlate with the experimental data we obtained from XPS. In this model, we insert the numerical value $z_{ks} \approx 2$ Å and $z_{kk} \approx 2.5$ Å.

The summary of the intensity (proportional to $C_{\text{Sn,k}} n_{\text{Sn,k}}$) is given in the table below for two different $\lambda$ values.

| $I_{\text{Sn},S_s}$ | $I_{\text{Sn},S_{kk}}$ | $I_{\text{Sn},S_k}$ | $\lambda$ (Å) |
|---|---|---|---|
| 2.15 | 1.11 | 1.29 | 1 |
| 2.575 | 1.56 | 1.95 | 2 |

| $I_{\text{Sn},S_s}/I_{\text{Sn},S_{kk}}$ | $I_{\text{Sn},S_s}/I_{\text{Sn},S_k}$ | $I_{\text{Sn},S_k}/I_{\text{Sn},S_{kk}}$ | $\lambda$ (Å) |
|---|---|---|---|
| 1.94 | 1.67 | 1.16 | 1 |
| 1.65 | 1.32 | 1.25 | 2 |

Comparing with the experimental result of $I_{bright}/I_{dark} = 1.84 \pm 0.57$ from figure 2 in the main text, while we cannot tell if the darker region is coming from the single or the bi-layer kagome, we can safely infer that the bright regions have stanene termination.

### C. Micro-focused Laser Angle Resolved Photoemission Spectroscopy (µ-ARPES)

For the µ-ARPES measurements, we use a 6.01 eV 4[th]-harmonic generation continuous laser from LEOS as photon source, a custom-built micro-focusing lens to focus the beam spot diameter to $\sim 3$ µm, and an MB-Scientific analyzer equipped with a deflection angle mode to map the dispersion relation while keeping the area of interest intact (*i.e.*, not changing due to sample rotation). Typical energy and angular resolution were 3 meV / 0.2°. The pressure during the measurement is kept at $< 10^{-10}$ mbar. The sample is mounted on a conventional 6-axes ARPES manipulator described in Ref. [2] and sample position is scanned with an xyz stage of 100 nm resolution and $< 1$ µm bidirectional reproducibility. A more detailed explanation can be found in Ref. [3]. During the measurement, the temperature is first lowered to 6 K, where the sample is cleaved *in situ*, and the sample drift at subsequent higher temperatures is tracked by using the edges of the sample as reference. The samples are pre-aligned to the high symmetry cut with low energy electron diffraction (LEED) after cleaving.

The perpendicular momentum $k_z$ of the electrons measured by ARPES can be obtained from the expression

$$k_z = \sqrt{\frac{2m_e^*}{\hbar^2} * (K_{out} + V_o) - \frac{2m_e}{\hbar^2} K_{out} \sin^2 \phi}$$

where $m_e^*$ is the effective mass of the electron, $K_{out} = h\nu - w - |E_b|$ is the kinetic energy of the electron, $h\nu$ is the photon energy, $|E_b|$ is the binding energy of the electron, $w$ is the work function of the detector, $V_o$ is the inner potential of the material, and $\phi$ is the analyzer slit angle (more details can be found in these references [4,5]). The laser photon energy used, 6.01 eV, corresponds to a perpendicular momentum close to the center of the Brillouin zone $k_z \approx m \times \frac{2\pi}{c}$, $m_{6.01} \approx 5.93 \approx 6$, as shown in Fig. SI3, whose inner potential used is taken from Ref. [6]. The synchrotron photon energy used is 48 eV, which also lies roughly in the same position at the center of the Brillouin zone, $m_{48} \approx 11.98 \approx 12$. From Fig. SI3 we can see that those two photon energies differ by two Brillouin zones.

We can also estimate the uncertainty in $k_z$ determination ($\delta k_z$) from the escape depth of the electron. In this case, we can focus on the electron close to the Fermi energy ($K_{out} \approx 1.64$ eV), which has an escape depth around $\lambda \approx 60$ Å according to the universal curve [1], giving us $\delta k_z = \frac{1}{\lambda} \approx 0.016$ Å$^{-1} \approx 0.05$ ΓZ, as drawn in main figure 3 (b)(ii). We can estimate $\delta k_z$ from the experimental line widths using the following relation:

$$\delta k_z \approx \frac{\text{FWHM}_{\text{EDC,experiment}}}{\left(\frac{\partial E}{\partial k_z}\right)_{\text{DFT}}}$$

In our data, FWHM$_{\text{EDC,experiment}}$ from each band at the Fermi level is (see SI5 (a),(b))

$$F\text{WHM}_{\text{EDC},\alpha,E_F} \approx 0.05 \text{ eV}$$
$$F\text{WHM}_{\text{EDC},\gamma,E_F} \approx 0.012 \text{ eV}$$

The DFT calculation shows roughly each band has (see main figure 3(b)(ii))

$$\left(\frac{\partial E}{\partial k_z}\right)_{E_F} \approx 0.61 \frac{\text{eV}}{\Gamma Z}$$

These values give us an estimation of $\delta k_z$ ranging

$$\delta k_z \approx (0.02 - 0.08) \Gamma Z$$

agreeing (by the order) with the estimation from the universal curve.

**Synchrotron vs Laser ARPES**

Twinned domains exist in Fe$_3$Sn$_2$ and will both contribute to the ARPES data if the beam spot is bigger than the size of the domains. For example, we show in Fig. SI4a an example of ARPES data measured with 48 eV photon energy, temperature 17 K, and a spot size of ~35 $\mu m$ collected at the SIS beamline, Swiss Light Source, equipped with Scienta R4000 hemispherical analyser. The overall feature shows a 6-fold pattern with no indication of a 3-fold pattern. A closer look around the center also shows a rather blurred circular shape, which can be attributed to a combination of all twinned domains that it may cover. This central feature can in fact be reconstructed from the laser ARPES results by combining data from both twinned domains and different polarizations (LH+LV) to give us the picture shown in Fig. SI4b. Indeed, we reproduce the synchrotron ARPES data but with higher momentum resolution.

**Fitting momentum distribution curve (MDC)**

The momentum distribution curves (MDC) are fitted to obtain the half peak width (Δk) which can be related to the averaged scattering length of the quasiparticle as $\lambda = \frac{1}{\Delta k}$. We obtain this peak by fitting the MDC peak with a Voigt lineshape, *i.e.*, a Lorentzian convolved with a Gaussian, where the Gaussian simulates the instrument response. For momentum scans (MDC), the latter combines angular and energy resolutions to yield a full-width-at-half-maximum (FWHM)

$$\delta k \approx \sqrt{(\delta k_x)^2 + (\delta k_y)^2} = \sqrt{(\delta k_x)^2 + \left(\delta E \left(\frac{\partial E}{\partial k}\right)^{-1}\right)^2}$$

where $v_F = \frac{1}{\hbar}\frac{\partial E}{\partial k}\big|_F$. In this case we set $\delta k_x \approx 0.001$ Å$^{-1}$ and $\delta E \approx 3$ meV and $\frac{\partial E}{\partial k}$ as listed in table I in the main manuscript. We can therefore see that typically the broadening in the MDC is due to the broadening in the energy distribution curve (EDC) because $\delta k_x < \delta E \left(\frac{\partial E}{\partial k}\right)^{-1}$.

**Fitting sharp quasi particle peak ($\beta$) EDC**

Figure SI5(c) shows the sharp quasiparticle peak of $\beta$ at the Fermi level, fitted with a Voigt function to represent the detector response. We obtain the resolution corrected Lorentzian FWHM of 2.5 meV.

### D. Density Functional Theory Calculations

Electronic structure calculations are performed within density functional theory (DFT) using the projector augmented wave method (PAW) [7] as implemented in the Viena *ab initio* simulation package (VASP) [8,9]. The exchange-correlation potential is described using the Perdew-Burke-Ernzerhof (PBE) [10] functional within the generalized-gradient approximation (GGA). Plane waves are used as a basis set with a kinetic energy cutoff of 450 eV. To sample the Brillouin zone, we employ a 24 × 24 × 24 Monkhorst-Pack grid. The spin-orbit coupling (SOC) is included in our calculations as described in Ref. [11]. We take into account the *p* and *d* semi-core states for the Fe and Sn PAW datasets used, respectively, and the simplified on-site Hubbard *U* correction [12] is added for the Fe *d* orbitals. We determined a Hubbard *U* value of 1.3 eV to reproduce the ARPES band structure around the Γ point at the Fermi level. Band structures calculated using several different *U* values with magnetic moments pointing in the kagomé plane are shown in Fig. SI7 (notation and labeling of Brillouin zone points are shown in Fig. SI6). In Fig. SI8, the band structure evolution as a function of magnetic moment direction ($M \parallel x$, $\theta = 70°$, and $M \parallel z$) is also plotted with selected *U* values displayed (0, 0.5 and 1.3 eV). Fig. SI9 simulates the band at ARPES energy window by plotting bands in the range of $k_z = 0 - 0.05$ ΓZ.

**"Dirac points" in DFT calculations at various U values**

We attempted to reproduce the "Dirac points" predicted in the tight-binding model by changing the *U* value in the DFT calculations as shown in Fig. SI10. Our results show that the "Dirac points" are not reproduced for *U* values between 0 and 1.3 eV. One may see that the closest resemblance of the band dispersion to the linear crossing is achieved at a *U* value of 0.5 eV or above, where the gap is narrow but never fully closes.

**References for the SI**

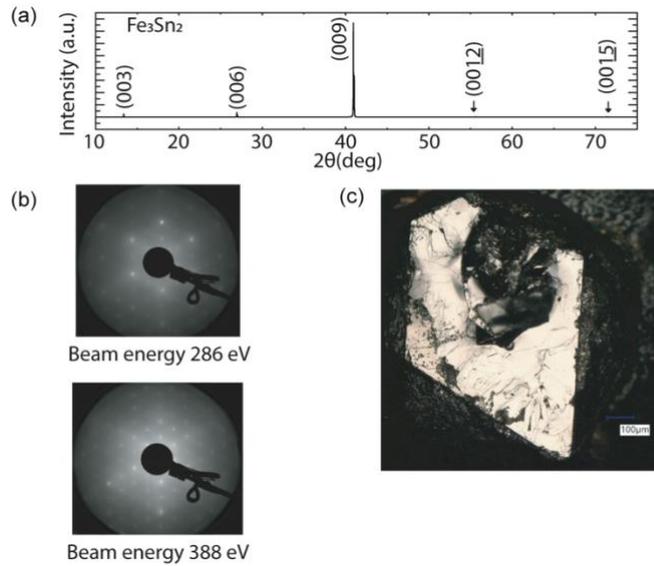

Fig. SI1 a. Powder diffraction pattern of polycrystalline Fe$_3$Sn$_2$ precursor, showing a good agreement with the previously reported $R\bar{3}m$ structure, with crystal axes $a, b = 5.34$ Å, $c = 19.80$ Å, and $\gamma = 120°$. b. Low energy electron diffraction of the cleaved surface of an Fe$_3$Sn$_2$ flake showing the correct symmetry pattern exposing (001) surface or *ab*-plane in conventional lattice. c. Typical cleaved surface picture of Fe$_3$Sn$_2$ showing a relatively flat surface suitable for ARPES measurement.

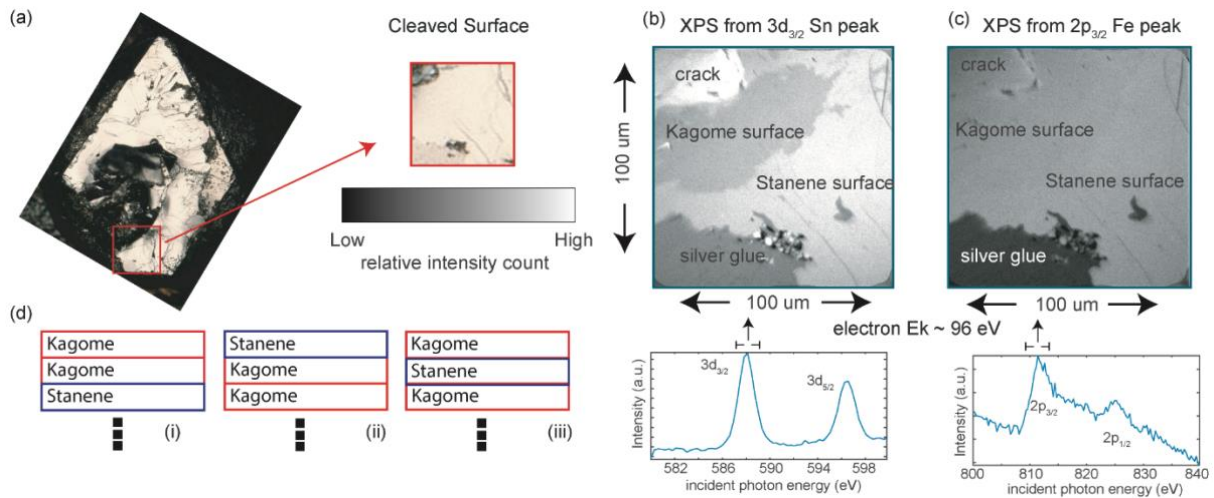

Fig. SI2 a. Optical image of a cleaved surface of $Fe_3Sn_2$ single crystal with highlighted area where the spatially resolved XPS was taken showing the area with crack and silver glue residue. b. Real space XPS intensity distribution of the Sn $3d_{3/2}$ photoemission line with the typical XPS spectrum shown below it. c. Real space XPS intensity distribution of the Fe $2p_{3/2}$ photoemission line with the typical XPS spectrum shown below it. d. The three possible surface terminations.

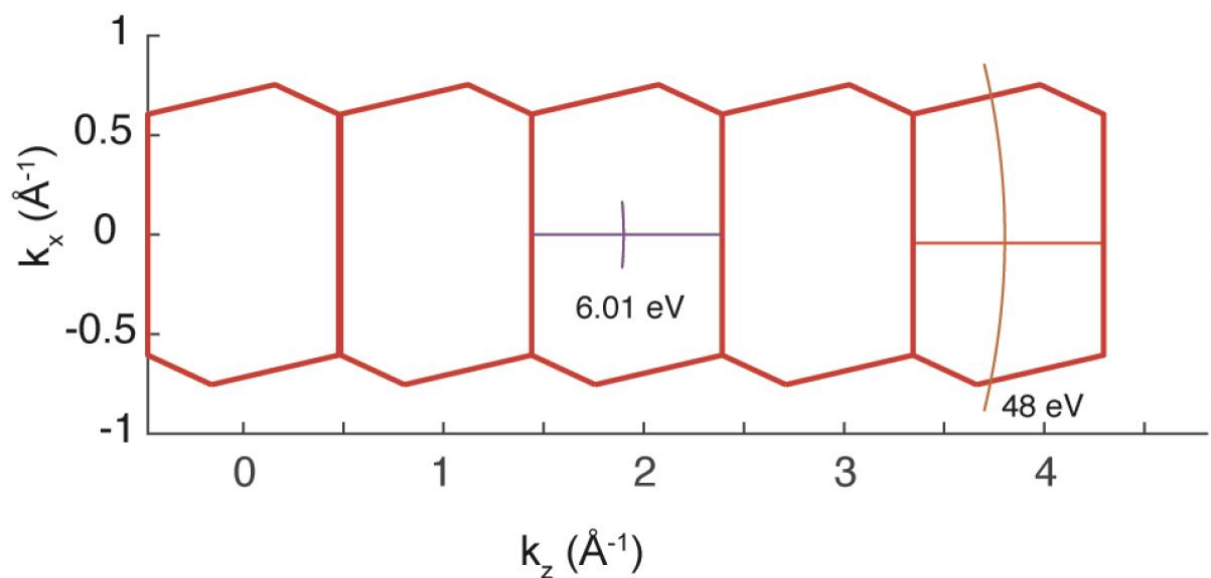

Fig. SI3 Perpendicular momentum ($k_z$) position for 6.01 eV laser energy and 48 eV synchrotron photon energy, showing that the ARPES data were collected on planes with $k_z \approx m \times \frac{2\pi}{c}$.

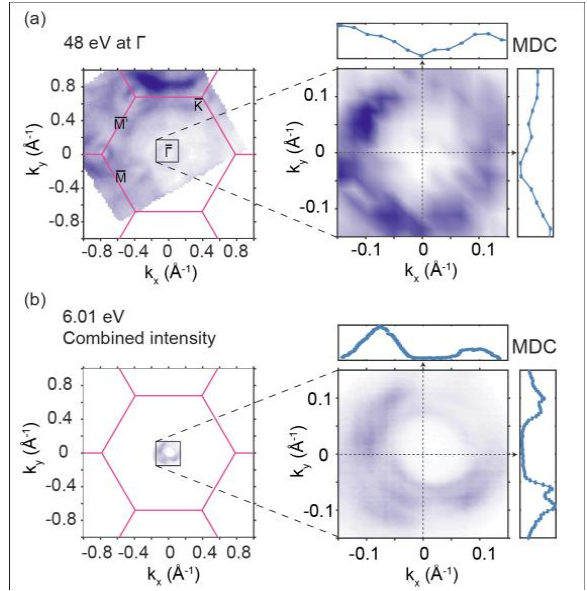

Fig. SI4 a. Fermi surface of $Fe_3Sn_2$ at $k_z \approx 0$ for ARPES data using synchrotron light (hν = 48 eV). b. Combined Fermi surface (LH + LV polarization) at $k_z \approx 0$ from twinned-domain (two domains combined together) laser ARPES data, 6.01 eV, simulating the synchrotron data probed with a larger beam spot.

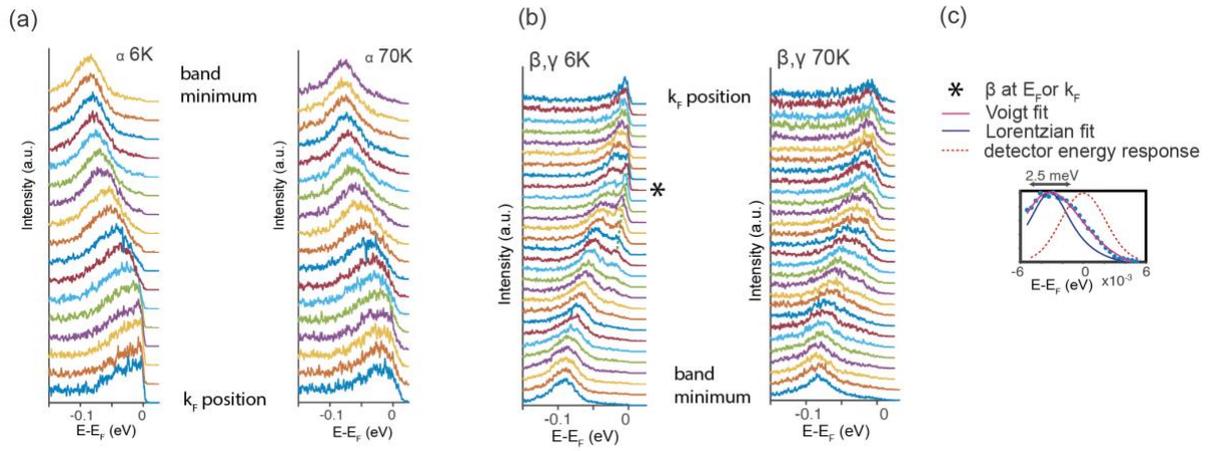

Fig. SI5 (a) and (b) Energy distribution curve (EDC) of $\alpha$ (a) and $\beta\,\&\,\gamma$ (b) at 6 K and 70 K. (c) Fitting of quasiparticle peak right at the $E_F$ (fitting the energy distribution curve or EDC) by convolving Lorentzian peak and gaussian peak (detector response).

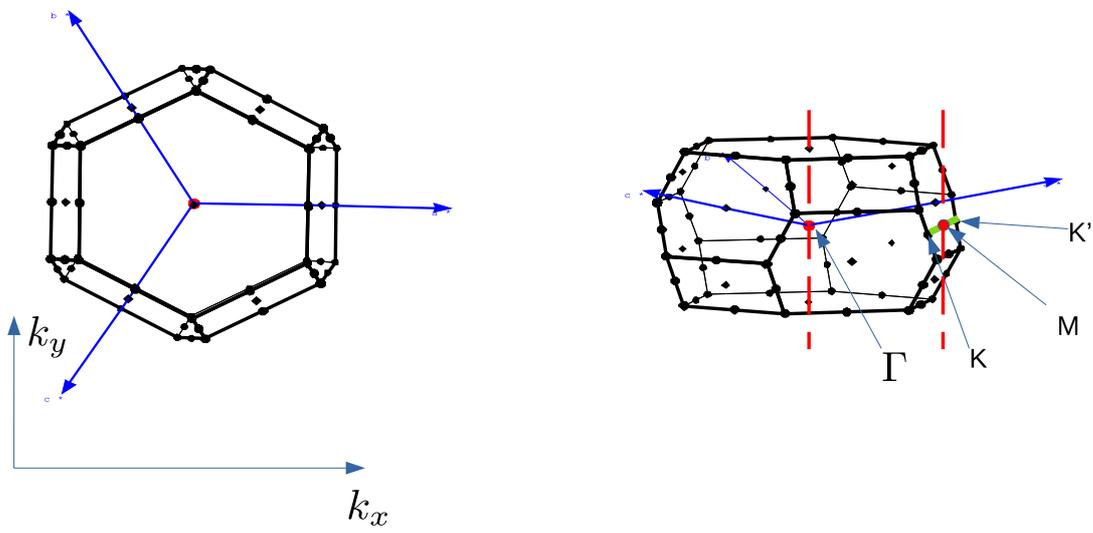

Fig. SI6 Brillouin zone of Fe$_3$Sn$_2$ and the naming convention for this work.

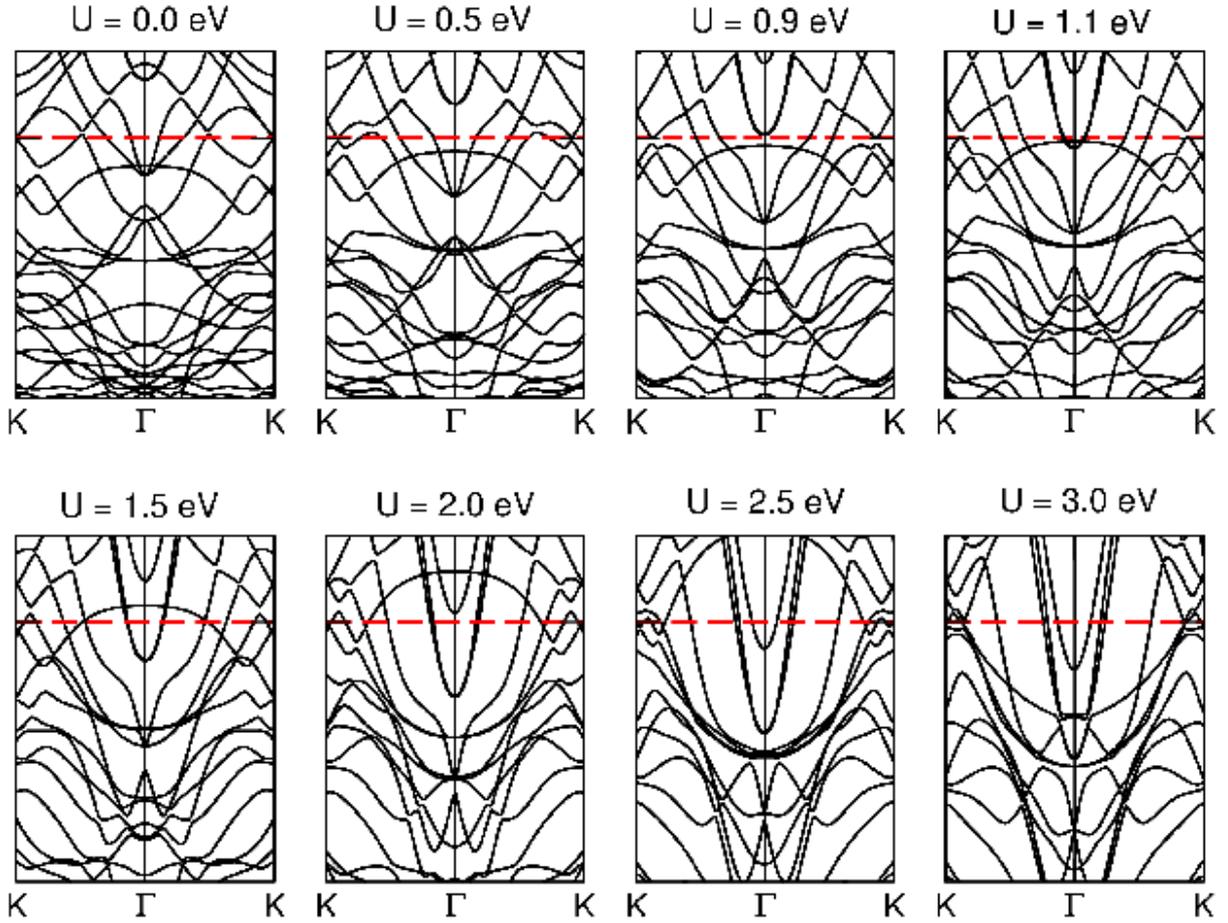
Fig. SI7 Band structure plots of Fe$_3$Sn$_2$ for $M \parallel x$ and various $U$ values in the direction ΓK, displaying the shift of bands, where the electron pocket close to the Fermi level is formed by different bands.

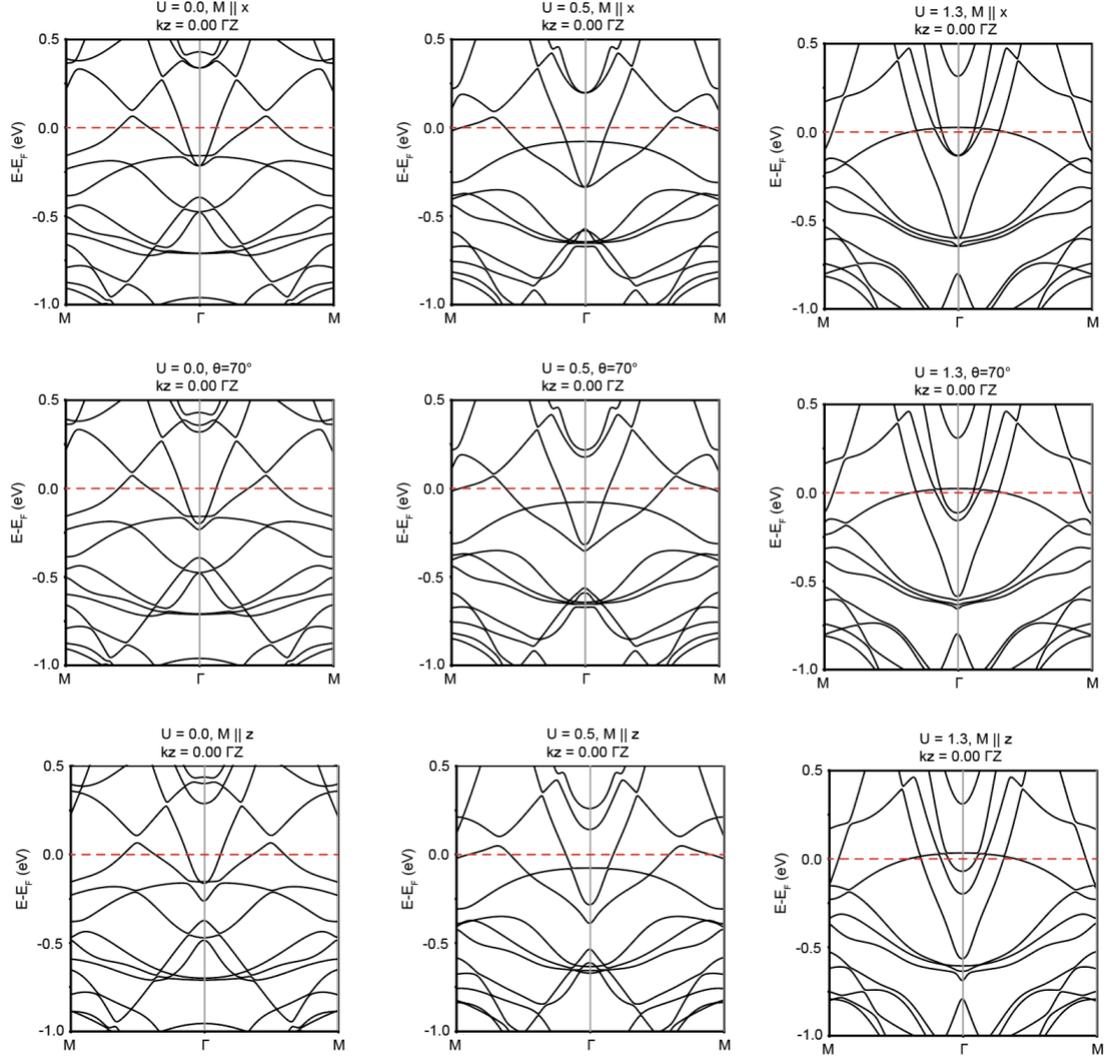

Fig. SI8 Band structure plots of Fe$_3$Sn$_2$ for selected $U$ values (0, 0.5, 1.3 eV) at various moment directions ($M \parallel x$, $\theta = 70°$, and $M \parallel z$) in the direction ΓM, displaying the shift of bands and band splitting as the moment points more towards the $c$-axis.

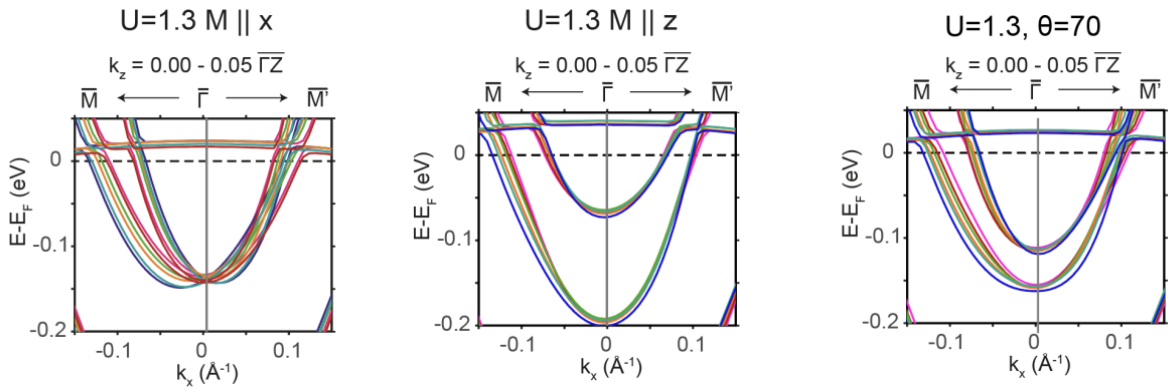

Fig. SI9 Band structure at U=1.3 eV plot around $k_z = 0 - 0.05$ ΓZ (to simulate $k_z$ broadening) for moment direction of $M \parallel x$, $M \parallel z$, and $\theta = 70°$.

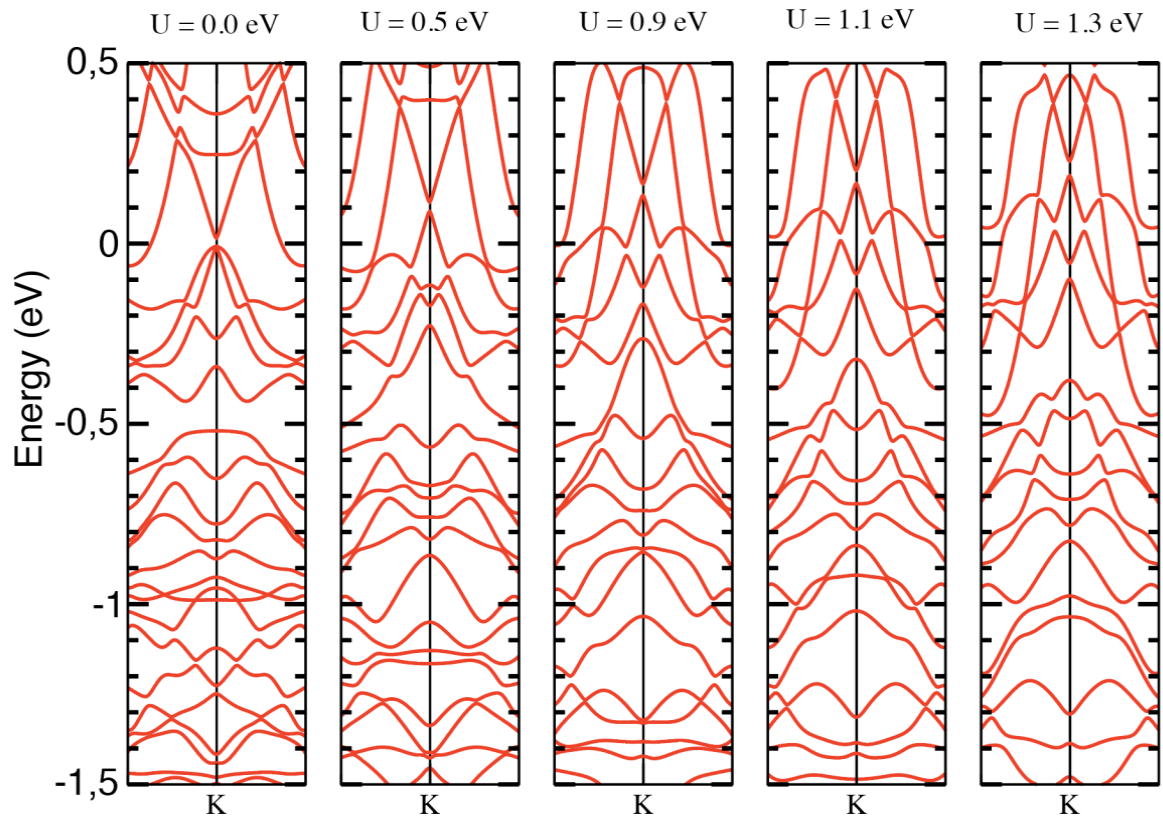

Fig. SI10 Band structure calculations of $Fe_3Sn_2$ performed with different $U$ values fail to reproduce the "Dirac points" at the $K$ point.